\documentclass[conference]{IEEEtran}
\ifCLASSINFOpdf
   \usepackage[pdftex]{graphicx}
\else
\fi
\usepackage{amsmath}
\usepackage{array}
\usepackage{booktabs}
\usepackage{multirow}
\usepackage{epsfig}
\begin{document}

\renewcommand\IEEEkeywordsname{Keywords}
%
\title{Study of Sparsity-Aware Set-Membership Adaptive Algorithms with Adjustable Penalties}

\author{\IEEEauthorblockN{Andre Flores}
\IEEEauthorblockA{Centre for Telecommunications Studies (CETUC)\\
PUC-Rio,\\
Rio de Janeiro, Brazil\\
Email: andre.flores@cetuc.puc-rio.br}
\and
\IEEEauthorblockN{Rodrigo C. de Lamare}
\IEEEauthorblockA{Centre for Telecommunications Studies (CETUC)\\
PUC-Rio,\\
Rio de Janeiro, Brazil\\
Department of Electronic Engineering, University of York, UK\\
Email: delamare@cetuc.puc-rio.br}}


%


\maketitle

\begin{abstract}
In this paper, we propose sparsity-aware data-selective adaptive
filtering algorithms with adjustable penalties. Prior work
incorporates a penalty function into the cost function used in the
optimization that originates the algorithms to improve their
performance by exploiting sparsity. However, the strength of the
penalty function is controlled by a scalar that is often a fixed
parameter. In contrast to prior work, we develop a framework to
derive algorithms that automatically adjust the penalty function
parameter and the step size to achieve a better performance.
Simulations for a system identification application show that the
proposed algorithms outperform in convergence speed
existing sparsity-aware algorithms.\\
\end{abstract}



%
\IEEEpeerreviewmaketitle

\section{Introduction}

A system is considered to be sparse if only a few of its elements
are nonzero values. A sparse signal can be represented as a vector
of a finite-dimensional space which can be expressed as a linear
combination of a small number of basis vectors of the related space.
There are many applications, such as echo cancellation, channel
equalization, and system identification, where sparse signals and
systems are found. However, traditional adaptive algorithms,
including the least-mean square (LMS), the affine projection (AP),
and the recursive least squares (RLS) do not exploit the sparsity of
the model \cite{Haykin2014}. When dealing with learning problems, we
attempt to extract as much as possible useful information from the
system to obtain better results. Under this scope, the sparsity of
systems has been the focus of many research works that are devoted
to improving the performance of adaptive algorithms.

One of the first approaches used to exploit sparsity was the
proportionate family of algorithms. These algorithms assign
proportional step sizes to different weights depending on their
magnitudes. These algorithms include the proportionate normalized
LMS (PNLMS) \cite{Duttweiler2000} and the improved PNLMS (IPNLMS)
\cite{BenestyGay2002}. 
Several versions of proportionate algorithms have been proposed such
as the $\mu$-law PNLMS (MPNLMS) \cite{DengDoroslovacki2006} and
improved MPNLMS (IMPNLMS) \cite{LiuFukamotoSaiki2008} algorithms. In
\cite{SousaTobiasSearaEtAl2010} an individual activation factor
PNLMS (IAF-PNLMS) algorithm was presented to better distribute the
adaptation over the coefficients. Additionally, the set-membership
NLMS (SM-NLMS)
\cite{GollamudiNagarajKapoor1998,smtvb,smce,smjio,sm-ccm,smbf,smcg}
and PNLMS (SM-PNLMS) \cite{WernerApolinarioDinizEtAl2005}, which is
a data-selective version of the PNLMS algorithm, has been developed.
The proportionate algorithms were also extended to the AP algorithm,
giving rise to the proportionate AP (PAP) and the improved PAP
(IPAP) \cite{HoshuyamaGoubranSugiyama2004} algorithms. The PAP
algorithm has also been discussed in
\cite{PaleologuCiochinaBenesty2010}. The main advantage of these
algorithms is that they accelerate the speed of convergence by
reusing multiple past inputs as a single input. Moreover, a
data-selective version, the set-membership PAP (SM-PAP) algorithm
has been introduced in \cite{WernerApolin/arioDiniz2007}.

In recent years, another approach to deal with sparsity based on
penalty functions has been adopted. In this context, a penalty
function is added to the cost function to take into account the
sparsity of the model and then a gradient-based algorithm is
derived. In \cite{ChenGuHero2009}, the zero-attracting LMS (ZA-LMS)
and the reweighted zero-attracting LMS (RZA-LMS) have been presented
and used for sparse system identification and other applications
\cite{aifir,jiolms,ccmmwf,wlmwf,wlrrbf,ccg,jiols,rccm,jiomimo,jidf,sjidf,gsc-radar,jio-radar,sa-stap,mserjidf,locsme,arh,als,jidfdoa,okspme,dce,damdc}
. This idea has been extended to the AP algorithm in
\cite{MengLamareNascimento2011}, where the zero-attracting AP
(ZA-AP) and the reweighted zero-attracting AP (RZA-AP) algorithms
have been proposed. Another example of this kind of algorithm is the
zero-attracting RLS (ZA-RLS) \cite{HongGaoChen2016}. Other versions
of the RLS algorithm that deal with sparsity in systems have been
studied in \cite{EksiogluTanc2011} and
\cite{AngelosanteBazerqueGiannakis2010}. There are also
data-selective versions of adaptive algorithms that incorporate a
penalty function \cite{LimaFerreiraMartinsEtAl2014}. A review of
common penalty functions used in the literature and another scheme
to treat sparsity has been reported in
\cite{LamareSampaio-Neto2014}.

In general, adaptive algorithms that use a penalty function are
computationally less expensive and they also achieve a better
trade-off between performance and complexity
\cite{DasAzpicueta-RuizChakrabortyEtAl2014} than proportionate
algorithms. However, a critical step in these algorithms is the
selection of the value of the regularization term. In this paper, we
propose a novel framework to derive data-selective algorithms with
adjustable penalties and develop algorithms to automatically adjust
the regularization term and the step-size. In particular, we devise
a framework for set-membership algorithms that can adjust the
step-size and the penalty based on the error bound. We then develop
sparsity-aware set-membership algorithms with adjustable penalties
using commonly employed penalty functions. Simulations show that the
proposed algorithms outperform prior art.

This paper is organized as follows. In Section II, the problem
formulation is presented. In Section III the proposed algorithms are
derived. Section IV presents the simulations and results of the
algorithms developed in an application involving system
identification. Finally, Section V presents the conclusions of this
work.

\section{Set-Membership Filtering and Problem Statement}

In set-membership filtering, the filter ${\bf w}(i)$ is designed to
achieve a specified bound on the magnitude of an estimate $y(i)$. As
a result of this constraint, set-membership adaptive algorithms will
only perform filter updates for certain data, resulting in
data-selective or sparse updates. Let $\Theta(i)$ represent the set
containing all possible ${\bf w}(i)$ that yield estimates upper
bounded in magnitude by an error bound $\gamma$. Thus, we can write
\begin{equation}
\Theta(i) = \bigcap_{{({\boldsymbol{x}}(i))\in \mathcal{S}}} \{ {\bf w} \in
{\mathcal{R}}^{M}:\mid y(i)\mid \leq \gamma \},
\end{equation}
where $\boldsymbol{x}(i)$ is the input vector, $\mathcal{S}$ is the set of
all possible data pairs $(d(i),~ \boldsymbol{x}\left(i\right)$ and the set
$\Theta(i)$ is referred to as the feasibility set, and any point in
it is a valid estimate $y(i)={\bf w}^{T}(i)\boldsymbol{x}\left(i\right)$.
Since it is not practical to predict all data pairs, adaptive
methods work with the membership set $\psi_{i}= \bigcap_{m=1}^{i}
{\mathcal{H}}_{m}$ provided by the observations, where
${\mathcal{H}}_{m}=\{{\bf w}(i) \in {\mathcal{R}}^{M}: |y(i)| \leq
\gamma\}$. In order to devise an effective set-membership algorithm,
the bound $\gamma$ must be appropriately chosen. Prior work has
considered data-selective or sparse updates, time-varying bounds
\cite{Lamare2009} and exploited sparsity in ${\bf w}(i)$. We review
the standard SM-NLMS algorithm next.

Let us consider the M-dimensional input vector expressed by
\begin{equation}
\boldsymbol{x}\left(i\right)=\left[\begin{array}{cccc}
x\left(i\right) & x\left(i-1\right) & \cdots & x\left(i-M+1\right)\end{array}\right]^{T}
\end{equation}
The output of the adaptive filter is given by
\begin{equation}
y\left(i\right)=\mathbf{w}^{T}\left(i\right)\boldsymbol{x}\left(i\right),
\end{equation}
and the error is computed as follows:
\begin{equation}
e\left(i\right)=d\left(i\right)-y\left(i\right)
\end{equation}
Let us consider a gradient descent approach, where our model is
updated by the recursive equation defined by
\begin{equation}
\mathbf{w}\left(i\right)=\mathbf{w}
\left(i-1\right)-\mu\left(i\right)\frac{\partial
J}{\partial{\mathbf{w}}\left(i-1\right)},\label{eq:gradient
descent}
\end{equation}
where $J$ is the cost function expressed by
\begin{equation}
J=\frac{1}{2}\mathbf{E}\left[|e(i)|^{2}\right]
\end{equation}
The gradient of $J$ is given by
\begin{equation}
\frac{\partial J}{\partial\mathbf{w}\left(i-1\right)}=-e\left(i\right)\boldsymbol{x}\left(i\right)
\end{equation}
Replacing this expression in the update equation leads to:
\begin{equation}
\mathbf{w}\left(i\right)=\mathbf{w}\left(i-1\right)+\mu\left(i\right) e\left(i\right)\boldsymbol{x}\left(i\right)
\end{equation}
An update of a set-membership algorithm takes place only if the
absolute value of the error exceeds the error bound so we have
\begin{align}
\gamma&=|d\left(i\right)-\mathbf{w}^{T}\left(i\right)\boldsymbol{x}\left(i\right)|\nonumber\\
&=|d\left(i\right)-\left(\mathbf{w}\left(i-1\right)+\mu\left(i\right) e\left(i\right)\boldsymbol{x}\left(i\right)\right)^{T}\boldsymbol{x}\left(i\right)\nonumber\\
&=|e\left(i\right)|\left(1-\mu\left(i\right)\parallel\boldsymbol{x}\left(i\right)\parallel^{2}\right),
\end{align}
which leads to the final step-size of the algorithm given by
\begin{equation}
\mu\left(i\right)=\begin{cases}
\frac{1}{\parallel\boldsymbol{x}\left(i\right)\parallel^{2}}\left(1-\frac{\gamma}{|e\left(i\right)|}\right), & |e\left(i\right)|>\gamma\\
0 & \mbox{otherwise}
\end{cases}\label{eq:step-size SMNLMS}
\end{equation}
resulting in the SM-NLMS update recursion:
\begin{equation}
\mathbf{w}\left(i\right)=\mathbf{w}\left(i-1\right)+\mu\left(i\right)e\left(i\right)\boldsymbol{x}\left(i\right),
\end{equation}
where $\mu\left(i\right)$ is given by \eqref{eq:step-size SMNLMS}.

Set-membership adaptive algorithms have sparse updates and variable
step-size, which are useful to ensure a fast learning. Prior work on
set-membership algorithms that exploit sparsity includes the studies
in
\cite{WernerApolinarioDinizEtAl2005,WernerApolin/arioDiniz2007,LimaFerreiraMartinsEtAl2014}.
However, the problem of adjusting the regularization term and the
resulting penalty imposed on the cost function remains open. In this
sense, we are interested in developing algorithms capable of
performing sparse updates and exploiting sparsity in signals and
systems. However, there has been no attempt to devise a strategy
based on the error bound to automatically adjust the regularization
term together with the step size. 

\section{Proposed Sparsity-Aware SM Algorithms with Adjustable Penalties}

In this section we introduce a framework for deriving sparsity-aware
set-membership adaptive algorithms with adjustable penalties using
arbitrary penalty functions. Then, we derive the proposed
sparsity-aware set-membership algorithms with adjustable penalties
based on a gradient descent approach.

\subsection{Derivation framework}

Let us consider a mean-square error cost function with a general
penalty function as described by
\begin{equation}
J\left[\mathbf{w}\left(i-1\right)\right]=\frac{1}{2}\mathbf{E}\left(|e\left(i\right)|^{2}\right)
+\alpha\left(i\right)f_{l}\left[\mathbf{w}\left(i-1\right)\right],
\end{equation}
where the function $f_{l}\left(\mathbf{w}\left[i-1\right]\right)$ is
a general penalty function used to improve the performance of
adaptive algorithms in the presence of sparsity and $\alpha(i)$ is a
regularization term that imposes the desired penalty.  The cost
function can rewritten as follows:
\begin{align}
J\left[\mathbf{w}\left(i-1\right)\right] & =\frac{1}{2}\mathbf{E}\left[|{d}\left(i\right)-\mathbf{w}^{T}\left(i-1\right)\boldsymbol{x}\left(i\right)|^{2}\right]\nonumber \\
 & \quad +\alpha\left(i\right)f_{l}\left[\mathbf{w}\left(i-1\right)\right]
\end{align}
Taking the instantaneous gradient of the cost function with respect
to $\mathbf{w}\left(i-1\right)$, we obtain
\begin{equation}
\frac{\partial J\left[\mathbf{w}\left(i-1\right)\right]}{\partial\mathbf{w}\left(i-1\right)}=-e\left(i\right)\boldsymbol{x}\left(i\right)+\alpha\left(i\right)\boldsymbol{p}_f\left(i\right),
\end{equation}
where we define $\boldsymbol{p}_f\left(i\right)=\frac{\partial f_{l}\left[\mathbf{w}\left(i-1\right)\right]}{\partial\mathbf{w}\left(i-1\right)}$. Replacing the result in the update equation, we get
\begin{equation}
\mathbf{w}\left(i\right)=\mathbf{w}\left(i-1\right)-\mu\left(i\right)\left(-e\left(i\right)\boldsymbol{x}\left(i\right)+\alpha\left(i\right)\boldsymbol{p}_f\left(i\right)\right).
\end{equation}

Note that we employ a time index in $\mu$ to designate a variable
step-size following the SM-NLMS approach and that the updates are
performed only if $|e\left(i\right)|>\gamma$, which leads to the
general equation to update the weights:
\begin{equation}
\mathbf{w}\left(i\right)=\mathbf{w}\left(i-1\right)+\mu\left(i\right)e\left(i\right)\boldsymbol{x}\left(i\right)-\rho\left(i\right)\boldsymbol{p}_f\left(i\right),\label{eq:weights update}
\end{equation}
where $\rho\left(i\right)=\mu\left(i\right)\alpha\left(i\right)$.
Using an equality constraint, i.e., the a posteriori error
$|e_{ap}\left(i\right)|=\gamma$ we obtain,
\begin{align}
\gamma&=|d\left(i\right)-\mathbf{w}^{T}\left(i\right)\boldsymbol{x}\left(i\right)|\\
 & =|d\left(i\right)-\mathbf{w}^{T}\left(i-1\right)\boldsymbol{x}\left(i\right)\nonumber \\
 & \quad -\left(\mu\left(i\right)e\left(i\right)\boldsymbol{x}\left(i\right)-\rho\left(i\right)\boldsymbol{p}_f\left(i\right)\right)^{T}\boldsymbol{x}\left(i\right)|.
\end{align}

Multiplying both sides of the last equation by $\frac{e_{ap}\left(i\right)}{|e_{ap}\left(i\right)|}$ results in
\begin{align}
\gamma\frac{e_{ap}\left(i\right)}{|e_{ap}\left(i\right)|}=&d\left(i\right)-\mathbf{w}^{T}\left(i-1\right)\boldsymbol{x}\left(i\right)\nonumber \\
 &-\mu\left(i\right)e\left(i\right)\parallel\boldsymbol{x}\left(i\right)\parallel^{2}+\rho\left(i\right)\left[\boldsymbol{p}_f\left(i\right)\right]^{T}\boldsymbol{x}\left(i\right)\\
\gamma\mbox{sign}\left(e_{ap}\left(i\right)\right)=&d\left(i\right)-\mathbf{w}^{T}\left(i-1\right)\boldsymbol{x}\left(i\right)\nonumber \\
 &-\mu\left(i\right)e\left(i\right)\parallel\boldsymbol{x}\left(i\right)\parallel^{2} +\rho\left(i\right)\left[\boldsymbol{p}_f\left(i\right)\right]^{T}\boldsymbol{x}\left(i\right)
\end{align}

Since the constraint forces that $|e_{ap}\left(i\right)|=\gamma$, then the function $\mbox{sign}\left(e_{ap}\left(i\right)\right)$ generates two possible equations given by,
\begin{align}
\gamma  &=e\left(i\right)-\mu\left(i\right)e\left(i\right)\parallel\boldsymbol{x}\left(i\right)\parallel^{2} +\rho\left(i\right)\left[\boldsymbol{p}_f\left(i\right)\right]^{T}\boldsymbol{x}\left(i\right)\label{eq:sign error ap plus}\\
-\gamma &=e\left(i\right)-\mu\left(i\right)e\left(i\right)\parallel\boldsymbol{x}\left(i\right)\parallel^{2} +\rho\left(i\right)\left[\boldsymbol{p}_f\left(i\right)\right]^{T}\boldsymbol{x}\left(i\right)\label{eq:sign error ap minus}
\end{align}
We can express equations \eqref{eq:sign error ap plus} and \eqref{eq:sign error ap minus} as a single equation as follows:
\begin{align}
e\left(i\right)\left(1-\frac{\gamma}{|e\left(i\right)|}\right) & =\mu\left(i\right)e\left(i\right)\parallel\boldsymbol{x}\left(i\right)\parallel^{2}\nonumber \\
 & \quad -\alpha\left(i\right)\mu\left(i\right)\left[\boldsymbol{p}_f\left(i\right)\right]^{T}\boldsymbol{x}\left(i\right),\label{eq:alfa mi variable}
\end{align}
where we take into account that the term $\left(1+\frac{\gamma}{|e\left(i\right)|}\right)$ would produce a growing step-size, leading to a divergent algorithm. Isolating the step-size from the last equation we obtain
\begin{equation}
\mu\left(i\right)=\frac{e\left(i\right)\left(1-\frac{\gamma}{|e\left(i\right)|}\right)}{\left(e\left(i\right)\parallel\boldsymbol{x}\left(i\right)\parallel^{2}-\alpha\left(i\right)\left[\frac{\partial f_{l}\left[\mathbf{w}\left(i-1\right)\right]}{\partial\mathbf{w}\left(i-1\right)}\right]^{T}\boldsymbol{x}\left[i\right]\right)}.\label{eq:mi}
\end{equation}
We then use equation \eqref{eq:alfa mi variable} to update $\alpha(i)$ as follows:
\begin{align}
\alpha\left(i+1\right)\mu\left(i\right)\left[\boldsymbol{p}_f\left(i\right)\right]^{T}\boldsymbol{x}\left(i\right) & =e\left(i\right)\frac{\gamma}{|e\left(i\right)|}\nonumber \\
 & 
 +e\left(i\right)\mu\left(i\right)\parallel\boldsymbol{x}\left(i\right)\parallel^{2}-e\left(i\right),
\end{align}
\begin{equation}
\alpha\left(i+1\right)=\frac{e\left(i\right)\left[\frac{\gamma}{|e\left(i\right)|}+\mu\left(i\right)\parallel\boldsymbol{x}\left(i\right)\parallel^{2}-1\right]}{\mu\left(i\right)\left[\boldsymbol{p}_f\left(i\right)\right]^{T}\boldsymbol{x}\left(i\right)}\label{eq:alpha}
\end{equation}
Equations \eqref{eq:weights update}, \eqref{eq:mi}, \eqref{eq:alpha}
fully describe the proposed sparsity-aware SM-NLMS algorithm with
adjustable penalties. We can easily show that if we set $\alpha(i)$
to zero, which means that there is no penalty function being
applied, then we get the conventional step size of the SM-NLMS
algorithm. Table \ref{tab:Penalty-functions} summarizes the penalty
functions used and their derivatives.

\begin{table}
\caption{\label{tab:Penalty-functions}Penalty Functions}
\begin{centering}
\begin{tabular}{ll}
\toprule \textbf{Function} & \textbf{Partial
derivative}\tabularnewline \midrule
$f_{l}\left[\mathbf{w}\left(i\right)\right]=\parallel\mathbf{w}\left(i\right)\parallel_{1}$
& $\mbox{sign}\left[\mathbf{w}\left(i\right)\right]$\tabularnewline
\midrule
$f_{l}\left[\mathbf{w}\left(i\right)\right]=\sum_{m=1}^{M}\log\left(1+\frac{|\mathrm{w}_m\left(i\right)|}{\varepsilon'}\right)$
&
$\left(\frac{\mbox{sign}\left[\mathbf{w}\left(i\right)\right]}{\varepsilon'+|\mathbf{w}\left(i\right)|}\right)$\tabularnewline
\midrule
$f_{l}\left[\mathbf{w}\left(i\right)\right]=\parallel\mathbf{w}\left(i\right)\parallel_{0}$
& $\beta
e^{-\beta|\mathbf{w}\left(i\right)|}\left[\mbox{sign}\left[\mathbf{w}\left(i\right)\right]\right]$\tabularnewline
$\quad~~~~~~~~~~~\approx \sum_{m=1}^{M} (1 - e^{-\beta |\mathrm{w}_m(i)|})$ \tabularnewline
\bottomrule
\end{tabular}
\par\end{centering}

\end{table}

\subsection{Proposed ZA-SM-NLMS-ADP algorithm}

In this section, we employ the previous derivation framework and the
$f_{l}\left[\mathbf{w}\left(i\right)\right]=\parallel\mathbf{w}\left(i\right)\parallel_{1}$
penalty function to devise the proposed zero-attracting SM-NLMS with
adjustable penalties  algorithm (ZA-SM-NLMS-ADP) . Substituting the
$l_{1}$ regularization function and its derivative, we obtain the
recursion and the step-size:
\begin{equation}
\mathbf{w}\left(i\right)=\mathbf{w}\left(i-1\right)+\mu\left(i\right)e\left(i\right)\boldsymbol{x}\left(i\right)-\rho\left(i\right)\mbox{sign}\left[\mathbf{w}\left(i-1\right)\right],
\end{equation}
\begin{equation}
\mu\left(i\right)=\frac{e\left(i\right)\left(1-\frac{\gamma}{|e\left(i\right)|}\right)}{\left(e\left(i\right)\parallel\boldsymbol{x}\left(i\right)\parallel^{2}-\alpha\left(i\right)\mbox{sign}\left[\mathbf{w}^{T}\left(i-1\right)\right]\boldsymbol{x}\left(i\right)\right)}\label{eq:mi-1}
\end{equation}
The regularization parameter that applies the adjustable penalties
is given by
\begin{equation}
\alpha\left(i+1\right)=\frac{e\left(i\right)\left(\frac{\gamma}{|e\left(i\right)|}+\mu\left(i\right)\parallel\boldsymbol{x}\left(i\right)\parallel^{2}-1\right)}{\mu\left(i\right)\mbox{sign}\left[\mathbf{w}^{T}\left(i-1\right)\right]\boldsymbol{x}\left(i\right)}
\end{equation}

\subsection{Proposed RZA-SM-NLMS-ADP algorithm}

Here, we consider the derivation framework and use the log-sum
penalty function
$f_{l}\left[\mathbf{w}\left(i\right)\right]=\sum_{n=1}^{N}\log\left(1+\frac{\mathbf{w}\left(i\right)}{\varepsilon'}\right)$
to develop the reweighted zero-attracting SM-NLMS with adjustable
penalties (ZA-SM-NLMS-ADP) algorithm whose recursions are given by
\begin{equation}
\mathbf{w}\left(i\right)=\mathbf{w}\left(i-1\right)+\mu\left(i\right)e\left(i\right)\boldsymbol{x}\left(i\right)-\rho\left(i\right)\left(\frac{\mbox{sign}\left[\mathbf{w}\left(i-1\right)\right]}{\varepsilon'+|\mathbf{w}\left(i-1\right)|}\right)
\end{equation}
\begin{equation}
\mu\left(i\right)=\frac{e\left(i\right)\left(1-\frac{\gamma}{|e\left(i\right)|}\right)}{\left(e\left(i\right)\parallel\boldsymbol{x}\left(i\right)\parallel^{2}-\alpha\left(i\right)\left(\frac{\mbox{sign}\left[\mathbf{w}^{T}\left(i-1\right)\right]}{\varepsilon'+|\mathbf{w^{T}}\left(i-1\right)|}\right)\boldsymbol{x}\left(i\right)\right)}\label{eq:mi RZA}
\end{equation}
\begin{equation}
\alpha\left(i+1\right)=\frac{e\left(i\right)\left(\frac{\gamma}{|e\left(i\right)|}+\mu\left(i\right)\parallel\boldsymbol{x}\left(i\right)\parallel^{2}-1\right)}{\mu\left(i\right)\left(\frac{\mbox{sign}\left[\mathbf{w}^{T}\left(i-1\right)\right]}{\varepsilon'+|\mathbf{w^{T}}\left(i-1\right)|}\right)\boldsymbol{x}\left(i\right)}\label{eq:alfa RZA}
\end{equation}

\subsection{Proposed EZA-SM-NLMS-ADP algorithm}

Finally, we consider the derivation framework and an approximation
to the $l_{0}$ regularization function given by
$f_{l}\left[\mathbf{w}\left(i\right)\right]= \sum_{m=1}^{M} (1 -
e^{-\beta |w_m(i)|})$ to devise the exponential zero-attractor
SM-NLMS with adjustable penalties (EZA-SM-NLMS-ADP) algorithm.
Consider the vector ${\bf z}(i)$ defined by
\begin{equation}
\mathbf{z}\left(i\right)=\beta\rho\left(i\right)e^{-\beta|\mathbf{w}\left(i\right)|}.
\end{equation}
Then, the update equation, the step size and the regularization term are given by
\begin{equation}
\mathbf{w}\left(i\right)=\mathbf{w}\left(i-1\right)+\mu\left(i\right)e\left(i\right)
\boldsymbol{x}\left(i\right)- \mathbf{z}\left(i\right)
\left({\rm
sign}\left(\mathbf{w}\left(i\right)\right)\right),\label{eq:updateEZA}
\end{equation}
\begin{equation}
\mu\left(i\right)=\frac{e\left(i\right)\left(1-\frac{\gamma}{|e\left(i\right)|}\right)}{\left(e\left(i\right)
\parallel\boldsymbol{x}\left(i\right)\parallel^{2}-\mathbf{z}^{T}\left(i\right) 
\left({\rm
sign}\left(\mathbf{w}^{T}\left(i-1\right)\right)\right)\boldsymbol{x}\left(i\right)\right)},\label{eq:miEZA}
\end{equation}
\begin{equation}
\alpha\left(i+1\right)=\frac{e\left(i\right)\left(\frac{\gamma}{|e\left(i\right)|}+\mu\left(i\right)
\parallel\boldsymbol{x}\left(i\right)\parallel^{2}-1\right)}{\mu\left(i\right) 
\mathbf{z}^{T}\left(i\right)
\left({\rm sign}\left(\mathbf{w}^{T}\left(i-1\right)\right)\right)\boldsymbol{x}\left(i\right)}.\label{eq:alfa EZA}
\end{equation}

\section{Simulations}

In this section we asses the performance of the proposed algorithms
for a sparse system identification task. For this purpose we
consider a system modeled by a finite impulse response (FIR) filter
with 64 taps in three different scenarios. The first scenario
represents a sparse system where only four taps have values
different from zero. In the second case, a semi-sparse model with 32 equispaced nonzero coefficients is
considered. In the last scenario, we explore the
case where there is no sparsity in the system, so that all
taps contribute to calculate the output. The input signal follows a
Gaussian distribution with a signal to noise ratio of 20 dB. The desired signal is corrupted by white additive Gaussian noise with $\sigma_{n}=0.04$. The step-size for the NLMS and the PNLMS algorithms was set to $0.5$ and the error bound was fixed to $\gamma=\sqrt{5}\sigma_n$. A maximum value of $\alpha(i+1)=10^{-3}$ was set to maintain the stability of the algorithms.

\begin{figure}[ht]
\begin{center}
\def\epsfsize#1#2{0.95\columnwidth}
\epsfbox{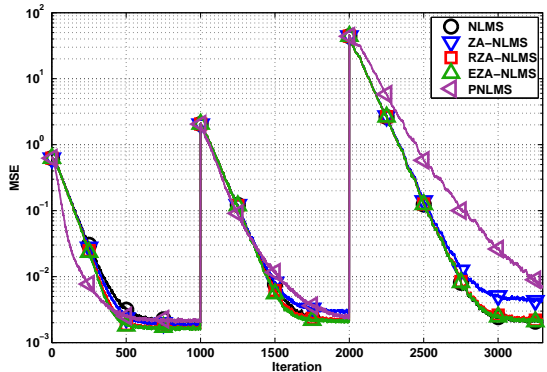}\caption{Learning curves of the NLMS-based
algorithms \label{fig:NLMS based algorithms}}
\end{center}
\end{figure}

In the first example, we compare the performance of NLMS-type
algorithms without adjustable penalties. Each algorithm runs for
3500 iterations, where the first 1000 corresponds to the first
scenario described, the next 1000 iterations corresponds to the
second scenario and the last 1500 iterations considered the third
scenario. A total of 3000 runs were performed and then averaged to
obtain the final learning curves.  The results shown in Fig.
\ref{fig:NLMS based algorithms} indicate that the sparsity-aware
SM-NLMS algorithms with different penalty functions outperform the
conventional NLMS and the PNLMS algorithms.

\begin{figure}[ht]
\begin{center}
\def\epsfsize#1#2{0.95\columnwidth}
\epsfbox{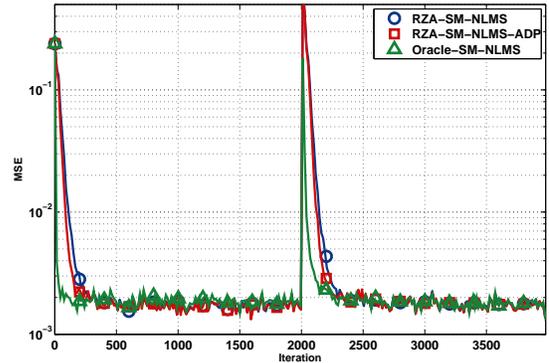}\caption{Learning curves of the RZA-SM-NLMS based
algorithms} \label{fig:Zeroattracting}
\end{center}
\end{figure}

In the second example, we evaluate the performance of the proposed
RZA-SM-NLMS-ADP algorithm. For this comparison we also considered
the oracle SM-NLMS algorithms \cite{LamareSampaio-Neto2014} that
assumes the knowledge of the positions of the nonzero coefficients.
In this sense, the oracle algorithm fully exploits the sparsity of
the system, being considered as the optimal algorithm. In this
example, a total of 4000 iterations were performed, where the first
2000 iterations corresponds to the sparse scenario and the last 2000
iterations considered the semi-sparse scenario. All other parameters
remain the same. The results depicted in Fig.
\ref{fig:Zeroattracting} show that the adjustable penalties
$\alpha(i)$ can provide a small but consistent gain over the fixed
penalty approach. Table \ref{tab:Updates} summarizes the update rate
performed by the proposed algorithms in a sparse scenario.
\begin{table}
\caption{\label{tab:Updates}\% of updates}
\begin{centering}
\begin{tabular}{lc}
\toprule \textbf{Algorithm} & \textbf{Update Rate}\tabularnewline
\midrule ZA-SM-NLMS-ADP & $\%$\tabularnewline \midrule
RZA-SM-NLMS-ADP & $\%$\tabularnewline \midrule EZA-SM-NLMS-ADP

& $\%$\tabularnewline
\bottomrule
\end{tabular}
\par\end{centering}
\end{table}

\begin{figure}[ht]
\begin{center}
\def\epsfsize#1#2{0.95\columnwidth}
\epsfbox{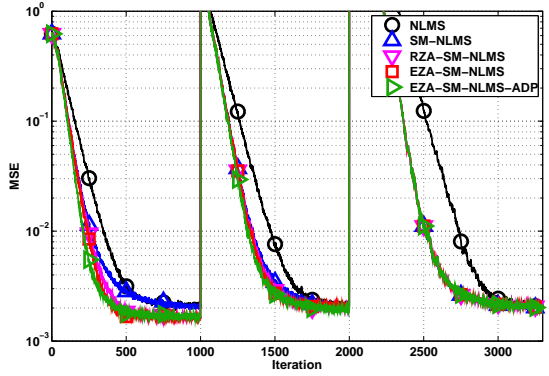}\caption{Learning curves of the proposed
EZA-SM-NLMS-ADP and other algorithms \label{fig:EZA}}
\end{center}
\end{figure}

In the third example, we assess the proposed EZA-SM-NLMS-ADP
algorithm against the other proposed and existing techniques. The
results shown in Fig. \ref{fig:EZA} demonstrate that EZA-SM-NLMS-ADP
has the fastest convergence speed among the conventional and
sparsity-aware algorithms.

\begin{figure}[ht]
\begin{center}
\def\epsfsize#1#2{0.95\columnwidth}
\epsfbox{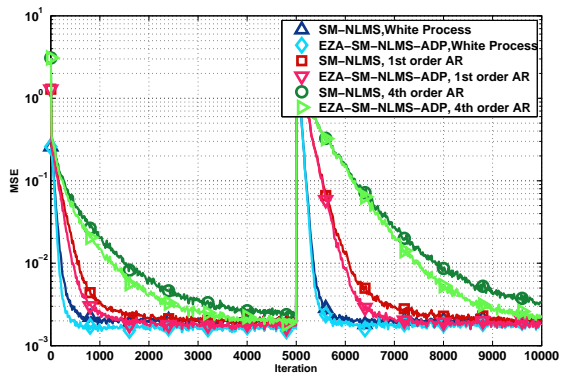}\caption{Performance of the EZA-SM-NLMS-ADP and the
SM-NLMS algorithm with correlated inputs\label{fig:EZAcorrelated}}
\end{center}
\end{figure}

Finally, we consider two different correlated inputs to evaluate the
performance of the proposed EZA-SM-NLMS-ADP algorithm. The input is
generated by a white Gaussian sequence $v\left(i\right)$,
uncorrelated with the noise. Then this signal is passed through two
different IIR filters described by
\begin{align}
x_1\left(i\right)=&0.7x\left(i-1\right)+v\left(i\right)\\
x_4\left(i\right)=&0.8x\left(i-1\right)+0.19x\left(i-2\right)+0.09x\left(i-3\right)\nonumber\\
&-0.5x\left(i-4\right)+v\left(i\right),
\end{align}
which corresponds to first- and fourth-order autoregressive (AR)
processes, respectively \cite{Haykin2014}. For the learning curves,
we consider a total of 5000 iterations, where the first 5000
iterations corresponds to the sparse scenario and the last set of
iterations represent the semi-sparse scenario. The results in Fig.
\ref{fig:EZAcorrelated} show that a correlated input slows the
converge speed and increases the steady-state MSE. In such cases,
applying a penalty function improves both results, the convergence
speed and the steady-state MSE.

\section{Conclusions}

In this paper data selective sparsity-aware algorithms with
adjustable penalty functions have been presented, namely, the
ZA-SM-NLMS-ADP, the RZA-SM-NLMS-ADP and the EZA-SM-NLMS-ADP adaptive
algorithms. These algorithms have a faster convergence speed than
conventional algorithms that implement fixed penalty functions. In
addition, the data-selective updates performed by these algorithms
can save computational resources. Future work will focus on the
statistical analysis of the proposed algorithms.

\section*{Acknowledgment}

The authors would like to thank the CNPq, and FAPERJ Brazilian agencies for funding.



%

\bibliographystyle{IEEEtran}{sorting}
\bibliography{adaptivefilters}
\end{document}